\documentclass[11pt]{article}
\usepackage{amsmath}
\usepackage{amsfonts}
\usepackage{latexsym}

\advance \textheight by \topskip
\advance \textheight by 1pt
\makeatother
\def\bea{\begin{eqnarray}}
\def\ena{\end{eqnarray}}
\def\non{\nonumber}
\newcommand{\qed}{\hbox{\rule[-2pt]{3pt}{6pt}}}

\def\Pf{{\rm Pf}}

\newtheorem{theorem}{Theorem}

\newtheorem{lemma}{Lemma}

\def\pf{{\it Proof.}\,}

\def\qed{$\Box$}

\title{Giambelli type formulae in the BKP hierarchy}

\author{
Yoko Shigyo\thanks{
e-mail: yoko.shigyo@gmail.com}\\
Department of Mathematics, Tsuda College,\\
Kodaira, Tokyo, 187-8577, Japan \\
\\
\\
\\
 \\
}
\date{}
\begin{document}
\maketitle

\begin{abstract}
In this paper, we study Giambelli type formula in the KP and the BKP hierarchies.
Any formal power series $\tau(x)$ can be expanded by the Schur functions.
It is known that $\tau(x)$ with $\tau(0)=1$ is a solution of the KP hierarchy if and only if the coefficients of this expansion satisfy Giambelli type formula.
It is proved by using Sato's theory of the KP hierarchy.
Here we give an alternative proof based on the previously established results on the equivalence of the addition formulae and the KP hierarchy without using Sato's theory.
This method of the proof can also be applied to the case of the BKP hierarchy.
\end{abstract}

\section{Introduction}
Let $\chi_{\lambda}(x),\,x=(x_1,x_2,\cdots)$ be the Schur function corresponding to a partition  $\lambda=(\lambda_1,\cdots,\lambda_l)$.
It is known that a formal power series $\tau(x)$ can be expanded in terms of the Schur functions as 
\bea
&&
\tau(x)=\sum_{\lambda}\xi_{\lambda}\chi_{\lambda}(x),\,\,\xi_{\lambda}=\chi_{\lambda}(\tilde{\partial})\tau(x)|_{x=0},
\label{eq:300}
\ena
where
\bea
&&
\tilde{\partial}=(\tilde{\partial}_1,\tilde{\partial}_2,\tilde{\partial}_3,\cdots),\,\,\tilde{\partial}_n=\frac{1}{n}\frac{\partial}{\partial x_n}.
\non
\ena
If $\tau(x)$ is a solution of the KP hierarchy with $\tau(0)=1$, then the coefficients $\{\xi_{\lambda}\}$ satisfy Giambelli type formula for any $\lambda$:
\bea
&&
\xi_{(k_1,\cdots,k_n|l_1,\cdots,l_n)}=\det(\xi_{(k_i|l_j)})_{1\le i,j\le n},
\label{eq:301}
\ena
where $(k_1,\cdots,k_n|l_1,\cdots,l_n)$ is the Frobenius notation of the partition $\lambda$.
In fact the converse is true.
Namely if (\ref{eq:301}) is satisfied for any partition, then the function $\tau(x)$ given by (\ref{eq:300}) with $\tau(0)=1$ is a solution of the KP hierarchy. 
This result had been proved using Sato's theory of the KP hierarchy \cite{EH1,SS1}.
Here we prove a similar result for the BKP hierarchy \cite{DJKM1}.
In the case of the BKP hierarchy we need the Schur's Q- functions instead of the Schur functions.

Let $Q_{\lambda}(x),\,x=(x_1,x_3,\cdots)$ be the Schur's Q-function corresponding to a strict partition $\lambda=(\lambda_1,\cdots,\lambda_l)$ (see 3.1 for the definition).
Then any formal power series $\tau(x),\,x=(x_1,x_3,\cdots)$ can be expanded using $\{Q_{\lambda}(x)\}$ (see appendix C):   
\bea
&&
\tau(x)=\sum_{\lambda}\xi_{\lambda}Q_{\lambda}\left(\frac{x}{2}\right), \,\,\xi_{\lambda}=Q_{\lambda}(\tilde{\partial})\tau(x)|_{x=0}.
\non
\ena
The main result of this paper is to prove that $\tau(x)$ with $\tau(0)=1$ is a solution of the BKP hierarchy if and only if the coefficients $\{\xi_{\lambda}\}$ satisfy Giambelli type formula:
\bea
&&
\xi_{\lambda}=2^{-n}\Pf\left(\xi_{(\lambda'_i,\lambda'_j)}\right)_{1\le i,j\le 2n},
\non
\ena
for any strict partition $\lambda=(\lambda_1,\cdots,\lambda_l)$ where $\lambda'=(\lambda'_1,\cdots,\lambda'_{2n})$ is the partition corresponding to $\lambda$ (see (\ref{eq:500})).
The notation $\Pf(a_{ij})_{1\le i,j\le 2n}$ denotes the Pfaffian of the determinant $A=(a_{ij})_{1\le i,j\le 2n}$.

Let us briefly explain how we prove the result.
We first give an alternative proof in case of the KP hierarchy.
We use the equivalence of the KP hierarchy and the addition formulae proved in \cite{S1,TT1}.
Since such equivalence is proved also for the BKP hierarchy \cite{S1}, a similar proof can be applied to the case of the BKP hierarchy.

Recently Giambelli and Jacobi-Trudi type formulae for the expansion coefficients for solutions of the KP or the modified KP (mKP) hierarchies attract much attention in the study of solvable lattice models \cite{AKLTZ}.
It is interesting to study  applications of our results to solvable lattice models \cite{K1}.

This paper consists of two sections and three appendices.
In section 2, we consider Giambelli type formula in the KP hierarchy.
We first review the KP hierarchy and the results related with the addition formulae in \cite{S1}.
Using them we give the proof of the equivalence of Giambelli type formulae for $\{\xi_{\lambda}\}$ and the KP hierarchy. 
We consider the BKP hierarchy in section 3.
We first introduce the Schur's Q-function in this section.
Then we review the BKP hierarchy and the results in \cite{S1}.
Finally we prove the main result in this paper. 
In Appendices A and B, necessary facts on fermions and the boson-fermion correspondence are given. 
In Appendix C, we prove that a formal power series of variables $(x_1,x_3,x_5,\cdots)$ can be expanded in terms of the Schur's Q-function.

\section{KP hierarchy}

\subsection{Partitions}
Let us begin by fixing notation which is used in this paper.

A sequence $\lambda=(\lambda_1,\lambda_2,\cdots,\lambda_n)$ of non-increasing non-negative integers is called a partition.
The non-zero $\lambda_i$ are called the parts of $\lambda$.
The number of parts is called the length of $\lambda$ and is denoted by $l(\lambda)$.
A partition $\lambda$ can also be written using the Frobenius notation (see \cite{Mac} for the precise definition):
\bea
&&
\lambda=(k_1,\cdots,k_r|l_1,\cdots,l_r),
\non
\ena
where $k_1>k_2>\cdots>k_r\ge 0$ and $l_1>l_2>\cdots>l_r\ge0$.

\vskip5mm

\subsection{The KP hierarchy}
Set
\bea
&&
[\alpha]=(\alpha,\frac{\alpha^2}{2},\frac{\alpha^3}{3},\cdots) ,\hskip5mm \xi(x,k)=\sum_{n=1}^{\infty}x_n k^n ,\hskip5mm x=(x_1,x_2,x_3,\cdots), \hskip5mm y=(y_1,y_2,y_3,\cdots).
\non
\ena
The KP hierarchy \cite{DJKM1} is a system of non-linear equations for a function $\tau(x)$ given by
\bea
&&
\oint e^{-2\xi(y,k)}\tau(x-y-[k^{-1}])\tau(x+y+[k^{-1}])\frac{dk}{2\pi i}=0,
\label{eq:1}
\ena 
where the integral means taking the coefficient of $k^{-1}$ in the expansion of the integrand in the series of $k$.

We have the addition formulae for the tau-function of the KP hierarchy  \cite{SS1}:
\bea
&&
\sum_{i=1}^{n+1}(-1)^{i-1} \zeta(x;\beta_1 , \dots , \beta_{n-1} , \alpha_i)\zeta(x;\alpha_1 , \dots , \hat{\alpha}_i , \dots , \alpha_{n+1})=0 , \hskip5mm n\ge2,
\label{eq:31}
\ena
where
\bea 
&&
\zeta(x;\alpha_1 , \dots , \alpha_{n})=\prod_{i<j}^n\alpha_{ij}\tau(x+[\alpha_1]+\dots+[\alpha_{n}]) ,\,\,\,\alpha_{ij}=\alpha_i-\alpha_j,
\non
\ena
and $\hat{\alpha}_i$ denotes to remove $\alpha_i$.

In the case of $n=2$, (\ref{eq:31}) becomes 
\bea
&&
\alpha_{12}\alpha_{34}\tau(x+[\alpha_1]+[\alpha_2])\tau(x+[\alpha_3]+[\alpha_4])
\non
\\
&&
-\alpha_{13}\alpha_{24}\tau(x+[\alpha_1]+[\alpha_3])\tau(x+[\alpha_2]+[\alpha_4])
\non
\\
&&
+\alpha_{14}\alpha_{23}\tau(x+[\alpha_1]+[\alpha_4])\tau(x+[\alpha_2]+[\alpha_3])=0,
\label{eq:400}
\ena
which is the simplest addition formula called the three term equation.

Using (\ref{eq:400}) we obtain the following determinant formula \cite{S1}:
\bea
&&
\frac{\prod_{i<j}^n\alpha_{ij}\beta_{ji}}{\prod_{i,j=1}^{n}(\beta_i-\alpha_j)}\tau(x+\sum_{i=1}^{n}[\beta_i]-\sum_{i=1}^{n}[\alpha_i])=\tau(x)^{-n+1}\det \left(\frac{\tau(x+[\beta_i]-[\alpha_j])}{\beta_i-\alpha_j}\right)_{1 \le i,j \le n}, 
\label{eq:5}
\ena
for $n\ge 2$ , where $\alpha_i$ and $\beta_j$ are parameters.

The following theorem is proved in \cite{TT1,S1}.
\begin{theorem}
Equation (\ref{eq:400}) is equivalent to the KP hierarchy.
\end{theorem}

\subsection{Giambelli type formula in the KP hierarchy} 

By Cauchy's identity (see \cite{Mac} p.63 (4.3)), any formal power series $\tau(x)$ can be expanded as
\bea
&&
\tau(x)=\sum_{\lambda}\xi_{\lambda}\chi_{\lambda}(x),\,\,\,\xi_{\lambda}\in \mathbb{C},
\label{eq:3}
\ena
where $\chi_{\lambda}(x)$ is the Schur function defined, for any partition $\lambda=(\lambda_1,\cdots,\lambda_n)$, as 
\bea
&&
\chi_{\lambda}(x)=\det\left(p_{\lambda_i-i+j}(x)\right)_{1\le i,j\le n},\,\,\,\,e^{\sum_{m=1}^{\infty}x_m z^m}=\sum_{k=0}^{\infty}p_k(x)z^k.
\non
\ena
The coefficients $\{\xi_{\lambda}\}$ can be written as a derivative of $\tau(x)$:
\bea
&&
\xi_{\lambda}=\chi_{\lambda}(\tilde{\partial})\tau(x)|_{x=0},\hskip5mm\tilde{\partial}=(\tilde{\partial}_1,\tilde{\partial}_2,\tilde{\partial}_3,\cdots),\,\,\,\tilde{\partial}_n=\frac{1}{n}\frac{\partial}{\partial x_n}.
\label{eq:2}
\ena

For convenience we extend the definition of $\chi_{\lambda}(x)$ to any sequence $\lambda=(k_1,\cdots,k_n|l_1,\cdots,l_n)$ of integers as follows.

We define $\chi_{(k_1,\cdots,k_n|l_1,\cdots,l_n)}=0$ if some $k_i$ and $l_j$ are negative, and $\chi_{\lambda}(x)$ is skew symmetric in $(k_1,\cdots,k_n)$ and $(l_1,\cdots,l_n)$ respectively. 

The following theorem is shown in \cite{EH1} using Sato's theory on the universal Grassmann manifold \cite{SS1}.
\begin{theorem}
A formal power series $\tau(x)$ with $\tau(0)=1$ is a solution of the KP hierarchy if and only if the coefficients $\{\xi_{\lambda}\}$ satisfy Giambelli type formula:
\bea
&&
\xi_{(k_1,\cdots,k_n|l_1,\cdots,l_n)}=\det(\xi_{(k_i|l_j)})_{1\le i,j\le n},
\label{eq:4}
\ena
for any partition $(k_1,\cdots,k_n|l_1,\cdots,l_n)$.
\end{theorem}

\noindent{\bf{Remark.}}\,\,\,
There is a freedom to multiply the tau function of the KP hierarchy by constants.
Thus, if $\tau(0)\neq0$, we can always normalize $\tau(x)$ as $\tau(0)=1$.

\vskip5mm
We prove this theorem directly from the equation (\ref{eq:1}) without using Sato's theory. 

\begin{lemma}
Suppose that parameters $\alpha_i$, $\beta_j$ satisfy $|\beta_i|<|\alpha_j|$ for any $i$ and $j$.
We have the following equation:
\bea
&&
\frac{\prod_{i<j}^n\alpha_{ij}\beta_{ji}}{\prod_{i,j=1}^{n}(\beta_i-\alpha_j)}e^{\sum_{i=1}^n \xi(x,\beta_i)-\sum_{i=1}^n \xi(x,\alpha_i)}
\non
\\
&&
=\sum_{k_i,l_j>0}(-1)^{l_1+\cdots+l_n}\chi_{(k_1,\cdots,k_n|l_1,\cdots,l_n)}(x)\alpha_1^{l_1}\cdots\alpha_n^{l_n}\beta_1^{k_1}\cdots\beta_n^{k_n}.
\label{eq:9}
\ena 
\end{lemma}

\vskip5mm
\noindent\pf\,\,
Consider the vertex operators \cite{DJKM1} defined by 
\bea
&&
X(k)=e^{\xi(x,k)}e^{-\xi(\tilde{\partial},k^{-1})},
\non
\\
&&
X^*(k)=e^{-\xi(x,k)}e^{\xi(\tilde{\partial},k^{-1})}.
\non
\ena
These vertex operators satisfy the following exchange relations:
\bea
&&
X^*(k)X^*(l)=(1-\frac{l}{k})e^{-\xi(x,k)-\xi(x,l)}e^{\xi(\tilde{\partial},k^{-1})+\xi(\tilde{\partial},l^{-1})},
\non
\\
&&
X(k)X(l)=(1-\frac{l}{k})e^{\xi(x,k)+\xi(x,l)}e^{\xi(-\tilde{\partial},k^{-1})-\xi(\tilde{\partial},l^{-1})},
\non
\\
&&
X^*(k)X(l)=(1-\frac{l}{k})^{-1}e^{-\xi(x,k)+\xi(x,l)}e^{\xi(\tilde{\partial},k^{-1})-\xi(\tilde{\partial},l^{-1})},
\non
\ena
where $|k|>|l|$.\\
Applying the vertex operators to $1$ we have 
\bea
&&
X^*(\alpha_1)\cdots X^*(\alpha_n)X(\beta_n)\cdots X(\beta_1)\cdot 1
\non
\\
&&
=\prod_{i<j}\left(1-\frac{\alpha_j}{\alpha_i}\right)\left(1-\frac{\beta_i}{\beta_j}\right)\prod_{i,j=1}^{n}\left(1-\frac{\beta_i}{\alpha_j}\right)^{-1}e^{\sum_{i=1}^n(\xi(x,\beta_i)-\xi(x,\alpha_i))}e^{\sum_{i=1}^n(-\xi(\tilde{\partial},\beta_i^{-1})+\xi(\tilde{\partial},\alpha_i^{-1}))}\cdot 1
\non
\\
&&
=(-1)^{n}\frac{\prod_{i<j}^n\alpha_{ij}\beta_{ji}}{\prod_{i,j=1}^{n}(\beta_i-\alpha_j)}\prod_{i=1}^n (\beta_i^{1-i}\alpha_i^i) e^{\sum_{i=1}^n \xi(x,\beta_i)-\sum_{i=1}^n \xi(x,\alpha_i)}.
\label{eq:7}
\ena

Notice that $1=\langle0|e^{H(x)}|0\rangle$ where $H(x)$ is given by (\ref{eq:200}).
By the boson-fermion correspondence we have

\bea
&&
X^*(\alpha_1)\cdots X^*(\alpha_n)X(\beta_n)\cdots X(\beta_1)\langle0|e^{H(x)}|0\rangle
\non
\\
&&
=\prod_{i=1}^n\alpha_i^{i-1}\beta_i^{1-i}\sum_{k_i,l_j\in\mathbb{Z}}\langle0|e^{H(x)}\psi_{l_1}^*\cdots\psi_{l_n}^*\psi_{k_n}\cdots\psi_{k_1}|0\rangle \alpha_1^{-l_1}\cdots\alpha_n^{-l_n}\beta_1^{k_1}\cdots\beta_n^{k_n}
\non
\\
&&
=\prod_{i=1}^n\alpha_i^{i}\beta_i^{1-i}\sum_{k_i,l_j\in\mathbb{Z}}\langle0|e^{H(x)}\psi_{-l_1-1}^*\cdots\psi_{-l_n-1}^*\psi_{k_n}\cdots\psi_{k_1}|0\rangle)\alpha_1^{l_1}\cdots\alpha_n^{l_n}\beta_1^{k_1}\cdots\beta_n^{k_n}
\label{eq:82}
\\
&&
=\prod_{i=1}^n\alpha_i^{i}\beta_i^{1-i}\sum_{k_i,l_j>0}(-1)^{l_1+\cdots+l_n+n}\chi_{(k_1,\cdots,k_n|l_1,\cdots,l_n)}(x)\alpha_1^{l_1}\cdots\alpha_n^{l_n}\beta_1^{k_1}\cdots\beta_n^{k_n}.
\label{eq:8}
\ena
In deriving (\ref{eq:8}) from (\ref{eq:82}) we use (2.4.11) in \cite{DJKM1}:
\bea
&&
\chi_{Y}(x)=(-1)^{m_1+\cdots+m_k}\langle0|e^{H(x)}\psi_{m_1}^*\cdots\psi_{m_k}^*\psi_{n_k}\cdots\psi_{n_1}|0\rangle,
\label{eq:600}
\ena
where $Y=(n_1,\cdots,n_k|-m_1-1,\cdots,-m_k-1)$ for $m_i<0\le n_j$.
This Shur function (\ref{eq:600}) is skew symmetric in $(m_1,\cdots,m_k)$ and $(n_1,\cdots,n_2)$ respectively. 
Thus it satisfies the property of the extended Schur function.
By (\ref{eq:7}) and (\ref{eq:8}), we obtain (\ref{eq:9}).\qed

\vskip5mm
\noindent{\it{Proof of Theorem 1.}}\\
Firstly we prove (\ref{eq:4}) for a solution $\tau(x)$ of the KP hierarchy with $\tau(0)=1$.

By Lemma 1 we have
\bea
&&
\frac{\prod_{i<j}^n\alpha_{ij}\beta_{ji}}{\prod_{i,j=1}^{n}(\beta_i-\alpha_j)}\tau(x+\sum_{i=1}^{n}[\beta_i]-\sum_{i=1}^{n}[\alpha_i])
\non
\\
&&
=\frac{\prod_{i<j}^n\alpha_{ij}\beta_{ji}}{\prod_{i,j=1}^{n}(\beta_i-\alpha_j)}e^{\sum_{i=1}^n\xi(\tilde{\partial},\beta_i)-\sum_{i=1}^n\xi(\tilde{\partial},\alpha_i)}\tau(x)
\non
\\
&&
=\sum(-1)^{l_1+\cdots+l_n}\chi_{(k_1,\cdots,k_n|l_1,\cdots,l_n)}(\tilde{\partial})\tau(x)\alpha_1^{l_1}\cdots\alpha_n^{l_n}\beta_1^{k_1}\cdots\beta_n^{k_n}.
\label{eq:21}
\ena

The $n=1$ case of (\ref{eq:21}) gives 
\bea
&&
\frac{\tau(x+[\beta]-[\alpha])}{\beta-\alpha}=\sum(-1)^l \chi_{(k|l)}(\tilde{\partial})\tau(x)\alpha^l\beta^k.
\label{eq:6}
\ena

Substituting (\ref{eq:21}) and (\ref{eq:6}) to (\ref{eq:5}) we get
\bea
&&
\sum(-1)^{l_1+\cdots+l_n}\chi_{(k_1,\cdots,k_n|l_1,\cdots,l_n)}(\tilde{\partial})\tau(x)\alpha_1^{l_1}\cdots\alpha_n^{l_n}\beta_1^{k_1}\cdots\beta_n^{k_n}
\non
\\
&&
=\tau(x)^{-n+1}\sum(-1)^{l_1+\cdots+l_n}\det\left(\chi_{(k_i|l_j)}(\tilde{\partial})\tau(x)\right)_{1\le i, j\le n}\alpha_1^{l_1}\cdots\alpha_n^{l_n}\beta_1^{k_1}\cdots\beta_n^{k_n}.
\non
\ena
Thus
\bea
&&
\chi_{(k_1,\cdots,k_n|l_1,\cdots,l_n)}(\tilde{\partial})\tau(x)=\tau(x)^{-n+1}\det\left(\chi_{(k_i|l_j)}(\tilde{\partial})\tau(x)\right)_{1\le i, j\le n}.
\label{eq:14}
\ena
Setting $x=0$ in (\ref{eq:14}) and using (\ref{eq:2}) we get (\ref{eq:4}).

\vskip5mm
Conversely we show that $\tau(x)$ given by (\ref{eq:3}) with $\tau(0)=1$ is a solution of the KP hierarchy if $\{\xi_{\lambda}\}$ satisfy (\ref{eq:4}).

For $\lambda=(k_1,\cdots,k_n|l_1,\cdots,l_n)$, substitute the expression $\xi_\lambda$ of (\ref{eq:2}) to (\ref{eq:4}) and we get
\bea
&&
\chi_{\lambda}(\tilde{\partial})\tau(x)|_{x=0}=\det\left(\chi_{(k_i|l_j)}(\tilde{\partial})\tau(x)|_{x=0}\right)_{1\le i, j\le n}.
\non
\ena 
For parameters $\alpha_i$, $\beta_j$, we have
\bea
&&
\sum_{k_i,l_j>0}(-1)^{l_1+\cdots+l_n}\chi_{\lambda}(\tilde{\partial})\tau(x)|_{x=0}\alpha_1^{l_1}\cdots\alpha_n^{l_n}\beta_1^{k_1}\cdots\beta_n^{k_n}
\non
\\
&&
=\sum_{k_i,l_j>0}(-1)^{l_1+\cdots+l_n}\det\left(\chi_{(k_i|l_j)}(\tilde{\partial})\tau(x)|_{x=0}\right)_{1\le i, j\le n}\alpha_1^{l_1}\cdots\alpha_n^{l_n}\beta_1^{k_1}\cdots\beta_n^{k_n}.
\label{eq:24}
\ena
Here we recall that (\ref{eq:21}) is valid for an arbitrary formal power series $\tau(x)$ not necessarily a solution of the KP hierarchy.
Then using (\ref{eq:21}) and (\ref{eq:6}), (\ref{eq:24}) becomes 
\bea
&&
\frac{\prod_{i<j}\alpha_{ij}\beta_{ji}}{\prod_{i,j=1}^{n}(\beta_i-\alpha_j)}\tau(\sum_{i=1}^{n}[\beta_i]-\sum_{i=1}^{n}[\alpha_i])=\det \left(\frac{\tau([\beta_i]-[\alpha_j])}{\beta_i-\alpha_j}\right)_{1 \le i,j \le n}.
\label{eq:25}
\ena
Let us consider the $\infty\times n$ matrix $A=\left(\frac{\tau([\beta_i]-[\alpha_j])}{\beta_i-\alpha_j}\right)_{1\le i,\,\,1\le j\le n}$ and the Pl\"{u}cker relations for minor determinants of this matrix.
They are
\bea
&&
\sum_{i=1}^{n+1}(-1)^{i-1} \Delta(K_i)\Delta(L_i)\tau(\sum_{r\in K_i}[\beta_r]-\sum_{s=1}^n[\alpha_s])\tau(\sum_{r\in L_i}[\beta_r]-\sum_{s=1}^n[\alpha_s])=0,
\label{eq:26}
\ena
where 
\bea
&&
K_i=(k_1,\cdots,k_{n-1},l_i),\,\,\,L_i=(l_1,\cdots,\hat{l}_i,\cdots,l_{n+1}),
\non
\\
&&
\Delta(k_1,\cdots,k_n)=\prod_{i<j}^n(\beta_{k_i}-\beta_{k_j}).
\non
\ena
Here we consider $K_i$ as $\{k_1,\cdots,k_{n-1},l_i\}$ and $L_i$ as $\{l_1,\cdots,\hat{l}_i,\cdots,l_{n+1}\}$ as sets if they appear in the summation symbols.

Set
\bea
&&
l_{n+1}=k_1,l_n=k_2,l_{n-1}=k_3,\cdots, l_4=k_{n-2},
\non
\ena
then we have
\bea
&&
\Delta(K_4)
=\Delta(K_5)
=\cdots
=\Delta(K_{n+1})=0.
\non
\ena
Setting $\tilde{x}=(\tilde{x}_1,\tilde{x}_2,\cdots)=-\sum_{i=1}^n[\alpha_i]$ we obtain
\bea
&&
\Delta(K_1)\Delta(L'_1)\tau(\tilde{x}+\sum_{i=1}^{n-1}[\beta_{k_i}]+[\beta_{l_1}])\tau(\tilde{x}+\sum_{i=1}^{n-2}[\beta_{k_i}]+[\beta_{l_2}]+[\beta_{l_3}])
\non
\\
&&
-\Delta(K_2)\Delta(L'_2)\tau(\tilde{x}+\sum_{i=1}^{n-1}[\beta_{k_i}]+[\beta_{l_2}])\tau(\tilde{x}+\sum_{i=1}^{n-2}[\beta_{k_i}]+[\beta_{l_1}]+[\beta_{l_3}])
\non
\\
&&
+\Delta(K_3)\Delta(L'_3)\tau(\tilde{x}+\sum_{i=1}^{n-1}[\beta_{k_i}]+[\beta_{l_3}])\tau(\tilde{x}+\sum_{i=1}^{n-2}[\beta_{k_i}]+[\beta_{l_1}]+[\beta_{l_2}])=0,
\label{eq:27}
\ena
where $L'_i=(l_1,\cdots,\hat{l}_i,\cdots,{l_3},{k_{n-2}},\cdots,{k_1})$.
Note that $-k\tilde{x}_k=\alpha_1^k+\cdots+\alpha_n^k$ is a power sum symmetric function and $\tilde{x}_1,\cdots,\tilde{x}_n$ are algebraically independent.
Since $n$ is arbitrary, (\ref{eq:27}) is valid if  $\tilde{x}$ is replaced by an indeterminate $x$ by a similar arguments in \cite{S1}.
Shifting $x$ to $x-\sum_{i=1}^{n-2}[\beta_{k_i}]$ we get
\bea
&&
\beta_{k_{n-1}l_1}\beta_{l_2l_3}\tau(x+ [\beta_{k_{n-1}}]+[\beta_{l_1}])\tau(x+ [\beta_{l_2}]+[\beta_{l_3}])
\non
\\
&&
-\beta_{k_{n-1}l_2}\beta_{l_1l_3}\tau(x+ [\beta_{k_{n-1}}]+[\beta_{l_2}])\tau(x+ [\beta_{l_1}]+[\beta_{l_3}])
\non
\\
&&
+\beta_{k_{n-1}l_3}\beta_{l_1l_2}\tau(x+ [\beta_{k_{n-1}}]+[\beta_{l_3}])\tau(x+ [\beta_{l_1}]+[\beta_{l_2}])=0,
\label{eq:28}
\ena
which is the three term equation (\ref{eq:400}).
By Theorem 1 (\ref{eq:28}) is equivalent to the KP hierarchy.
Thus Theorem 2 is proved.\qed

\section{The BKP hierarchy}

\subsection{Schur's Q-function}
Let us consider the Schur's Q-function before talking about the BKP hierarchy.
The Schur's Q- function is defined for strict partitions.
A partition $\lambda=(\lambda_1,\cdots,\lambda_l)$ is strict if parts of $\lambda$ are distinct and positive, namely $\lambda_1>\cdots\lambda_l>0$.

For a strict partition $\lambda=(\lambda_1,\cdots,\lambda_l)$ we set 
\bea
&&
\lambda'=
\begin{cases}
(\lambda_1,\cdots,\lambda_l,0),\,\,&\text{if $l$ is odd},\\
(\lambda_1,\cdots,\lambda_l),\,\,&\text{if $l$ is even}.
\end{cases}
\label{eq:500}
\ena
Then the length of $\lambda'$ is always even.
  
For a non-negative integer $r$ define the symmetric polynomial $q_r$ of $\alpha=(\alpha_1,\cdots,\alpha_N)$ by
\bea
&&
\sum_{r\ge 0}q_r t^r=\prod_{i=1}^N\frac{1+t\alpha_i}{1-t\alpha_i}.
\non
\ena
For $r>s\ge 0$, we set 
\bea
&&
Q^{sym}_{(r,s)}=q_r q_s +2\sum_{i=1}^s (-1)^i q_{r+i}q_{s-i}.
\non
\ena
If $r<s$, we define $Q^{sym}_{(r,s)}$ as
\bea
&&
Q^{sym}_{(r,s)}=-Q^{sym}_{(s,r)}.
\non
\ena
For any strict partitions $\lambda=(\lambda_1,\cdots,\lambda_l)$, the Schur's Q-function is defined by 
\bea
&&
Q^{sym}_{\lambda}=\Pf\left(Q^{sym}_{(\lambda'_i,\lambda'_j)}\right)_{1\le i,j\le 2n},
\non
\ena
where $\Pf(a_{ij})_{1\le i,j\le 2n}$ denotes the Pfaffian of $A=(a_{ij})_{1\le i,j\le 2n}$.
We set $x_i=(\alpha^i_1+\cdots+\alpha^i_N)/i,\,\,N\ge |\lambda|$.
It is known that $Q^{sym}_{\lambda}$ can uniquely be expressed as a polynomial of $x=(x_1,x_3,x_5,\cdots)$. 
We denote this polynomial by $Q_{\lambda}(x)$.
Then we have the relation
\bea
&&
Q^{sym}_{\lambda}(x)=Q_{\lambda}(x),\,\,\,x_i=\frac{\alpha^i_1+\cdots+\alpha^i_N}{i}.
\non
\ena



We can expand any formal power series $\tau(x)$ of $x=(x_1,x_3,x_5,\cdots)$ as follows  (see appendix C):
\bea
&&
\tau(x)=\sum_{\lambda}\xi_{\lambda}Q_{\lambda}(\frac{x}{2}),
\label{eq:51}
\\
&&
\xi_{\lambda}=2^{-l(\lambda)}Q_{\lambda}(\tilde{\partial})\tau(x)|_{x=0},
\label{eq:111}
\ena
where $\lambda$ runs over all strict partitions.

Let us extend the definition of $Q_{\lambda}(x)$ to any sequence $\lambda=(\lambda_1,\cdots,\lambda_l)$ of integers as follows.

We define $Q_{(\lambda_1,\cdots,\lambda_l)}(x)=0$ of some $\lambda_i$ is negative, and 
\bea
&&
Q_{(\lambda_1,\cdots,\lambda_l)}(x)=sgn\sigma Q_{(\lambda_{\sigma(1)},\cdots,\lambda_{\sigma(l)})}(x)
\non
\ena
if $(\lambda_{\sigma(1)},\cdots,\lambda_{\sigma(l)})$ is a strict partition for some permutation $\sigma$.

\subsection{The BKP hierarchy}

Set
\bea 
&&
[\alpha]_o=(\alpha,\frac{\alpha^3}{3},\frac{\alpha^5}{5},\cdots),\hskip5mm \tilde{\xi}(x,k)=\sum_{n=1}^{\infty}x_{2n-1}k^{2n-1},\hskip5mm x=(x_1,x_3,x_5,\cdots),\hskip5mm y=(y_1,y_3,y_5,\cdots).
\non
\ena
We define the BKP hierarchy \cite{DJKM1} by
\bea
&&
\oint e^{-2\tilde{\xi}(y,k)}\tau(x-y-2[k^{-1}]_o)\tau(x+y+2[k^{-1}]_o)\frac{dk}{2\pi i k}=\tau(x-y)\tau(x+y).
\non
\ena

Define the components of the skew symmetric matrices $A=(a_{ij})_{0\le i,j\le n}$ and $A'=(a_{ij})_{1\le i,j\le n}$ by
\bea
&&
a_{0,j}=\tau(x+2[\alpha_j]_o),\hskip5mm  \hskip1cm a_{i,j}=\frac{\alpha_{ij}}{\tilde{\alpha}_{ij}}\tau(x+2[\alpha_i]_o+2[\alpha_j]_o),
\non
\ena
where $\tilde{\alpha}_{ij}=\alpha_i+\alpha_j$.
The addition formulae of the BKP hierarchy \cite{S1} are, for $n$ odd,
\bea
&&
\tau(x+2\sum_{i=1}^{n}[\alpha_i]_o)=\tau(x)^{(-n+1)/2}A_{1,2,\cdots,n}\Pf A,
\label{eq:56}
\ena
and, for $n$ even,
\bea
&&
\tau(x+2\sum_{i=1}^{n}[\alpha_i]_o)=\tau(x)^{(-n+2)/2}A_{1,2,\cdots,n}\Pf A',
\label{eq:57}
\ena
where
\bea 
&&
A_{1\dots n}=\prod_{i<j}^{n} \frac{\tilde{\alpha}_{ij}}{\alpha_{ij}}.
\non
\ena
The following theorem is proved in \cite{S1}.

\begin{theorem}
The case $n=3$ of (\ref{eq:56})
\bea
&&
A_{123}^{-1}\tau(x)\tau(x+2\sum_{i=1}^{3}[\alpha_i]_o)=
\frac{\alpha_{23}}{\tilde{\alpha}_{23}}\tau(x+2[\alpha_1]_o)\tau(x+2[\alpha_2]_o+2[\alpha_3]_o)
\non
\\
&&
\hskip4.5cm
-\frac{\alpha_{13}}{\tilde{\alpha}_{13}}\tau(x+2[\alpha_2]_o)\tau(x+2[\alpha_1]_o+2[\alpha_3]_o)
\non
\\
&&
\hskip4.5cm
+\frac{\alpha_{12}}{\tilde{\alpha}_{12}}\tau(x+2[\alpha_3]_o)\tau(x+2[\alpha_1]_o+2[\alpha_2]_o)
\label{eq:401}
\ena
is equivalent to the BKP hierarchy.
Equation (\ref{eq:401}) is called the four term equation.
\end{theorem}



\subsection{Giambelli type formula in the BKP hierarchy}

Our main theorem is
\begin{theorem}
A formal power series $\tau(x)$ with $\tau(0)=1$ is a solution of the BKP hierarchy if and only if the coefficients $\{\xi_{\lambda}\}$ in (\ref{eq:51}) satisfy
\bea
&&
\xi_{\lambda}=2^{-n}\Pf\left(\xi_{(\lambda'_i,\lambda'_j)}\right)_{1\le i,j \le 2n},
\label{eq:52}
\ena
for any strict partition $\lambda$, where $\lambda'=(\lambda'_1,\cdots,\lambda'_{2n})$ is defined by (\ref{eq:500}).
\end{theorem}

\vskip5mm

\begin{lemma}
We have the following equation:
\bea
&&
A_{1\cdots 2n}^{-1} e^{\sum_{i=1}^{2n}\sum_{k:odd}x_k\alpha_i^k}=\sum_{\lambda_i>0} Q_{(\lambda_1\cdots,\lambda_{2n})}\left(\frac{x}{2}\right)\alpha_1^{\lambda_1}\cdots\alpha_{2n}^{\lambda_{2n}},
\label{eq:62}
\ena
where parameters $\alpha_1,\cdots,\alpha_{2n}$ satisfy $|\alpha_1|>\cdots>|\alpha_{2n}|$.
\end{lemma}

\noindent\pf\,
Let 
\bea
&&
X_B(k)=e^{\sum_{n:odd}x_n k^n} e^{-2\sum_{n:odd}\tilde{\partial}_n k^{-n}}.
\non
\ena
We apply this vertex operator to $1=\langle0|e^{H_B(x)}|0\rangle$.

First, the exchange rule of the vertex operators is 
\bea
&&
X_B(k)X_B(l)=\frac{k-l}{k+l}e^{\sum x_n k^n+\sum x_n l^n}e^{-2\sum\tilde{\partial}_n k^{-n}-2\sum\tilde{\partial}_n k^{-n}}.
\label{eq:59}
\ena
Then we have
\bea
&&
X_B(\alpha_1)\cdots X_B(\alpha_{2n})\cdot 1=\prod_{i<j}\frac{\alpha_{ij}}{\tilde{\alpha}_{ij}}e^{\sum_{i=1}^{2n}\sum_{k:odd}x_k\alpha_i^k}.
\label{eq:60}
\ena
By the boson-fermion correspondence, we have
\bea
&&
X_B(\alpha_1)\cdots X_B(\alpha_{2n})\langle0|e^{H_B(x)}|0\rangle=2^{n}\langle0|e^{H_B(x)}\phi(\alpha_1)\cdots\phi(\alpha_{2n})|0\rangle
\non
\\
&&
\hskip5.2cm=2^{n}\langle0|e^{H_B(x)}\phi_{\lambda_1}\cdots\phi_{\lambda_{2n}}|0\rangle\alpha_1^{\lambda_1}\cdots\alpha_{2n}^{\lambda_{2n}}
\label{eq:81}
\\
&&
\hskip5.2cm=\sum_{\lambda_i>0} Q_{(\lambda_1\cdots,\lambda_{2n})}\left(\frac{x}{2}\right)\alpha_1^{\lambda_1}\cdots\alpha_{2n}^{\lambda_{2n}},
\label{eq:61}
\ena
where $H_B(x)$ is defined in (\ref{eq:201}). 
We get (\ref{eq:61}) from (\ref{eq:81}) using  (1.1.24) of \cite{O2} (see also \cite{DJKM1,Y1}):
\bea
&&
\langle0|e^{H_B(t)}\phi_{n_1}\cdots\phi_{n_k}|0\rangle=2^{-k/2}Q_{\lambda}\left(\frac{t}{2}\right),
\label{eq:601}
\ena
where $\lambda=(n_1,\cdots,n_k)$.
This function (\ref{eq:601}) is skew symmetric.
Thus it satisfies the property of the extended Schur's Q-function.
By (\ref{eq:60}) and (\ref{eq:61}), we have (\ref{eq:62}).\qed

\vskip5mm
\noindent{\it{Proof of Theorem 2.}}\,
The way to prove this theorem is similar to the case of the KP hierarchy.
We shall prove that $\{\xi_{\lambda}\}$ satisfy (\ref{eq:52}) if $\tau(x)$ is a solution of BKP heirarchy.
We consider (\ref{eq:57}) with $n$ replaced by $2n$, $n\ge2$.

By Lemma 2 we have
\bea
&&
A_{1\cdots 2n}^{-1}\tau(x+2\sum_{i=1}^{2n}[\alpha_i]_o)=A_{1\cdots 2n}^{-1}e^{\sum_{i=1}^{2n}\tilde{\xi}(2\tilde{\partial}_o,\alpha_i)}\tau(x)
\non
\\
&&
\hskip3.7cm=\sum Q_{(\lambda_1\cdots,\lambda_{2n})}(\tilde{\partial}_o)\tau(x)\alpha_1^{\lambda_1}\cdots\alpha_{2n}^{\lambda_{2n}},
\label{eq:63}
\ena
where $\tilde{\partial}_o=(\tilde{\partial}_1,\tilde{\partial}_3,\tilde{\partial}_5,\cdots)$.

The case of $n=1$ in (\ref{eq:63}) is 
\bea
&&
\frac{\alpha_{ij}}{\tilde{\alpha}_{ij}}\tau(x+2[\alpha_i]_o+2[\alpha_j]_o)=\sum Q_{(\lambda_i,\lambda_j)}(\tilde{\partial}_o)\tau(x)\alpha_i^{\lambda_i}\alpha_j^{\lambda_j}.
\label{eq:64}
\ena
Substituting (\ref{eq:63}) and (\ref{eq:64}) to (\ref{eq:57}), we have
\bea
&&
\sum Q_{(\lambda_1,\cdots,\lambda_{2n})}(\tilde{\partial}_o)\tau(x)\alpha_1^{\lambda_1}\cdots\alpha_{2n}^{\lambda_{2n}}
\non
\\
&&
=\tau(x)^{-n+1}\Pf\left(\sum Q_{(\lambda_i,\lambda_j)}(\tilde{\partial}o)\tau(x)\alpha_i^{\lambda_i}\alpha_j^{\lambda_j}\right)_{1\le i, j\le2n}
\non
\\
&&
=\tau(x)^{-n+1}\sum\Pf\left(Q_{(\lambda_i,\lambda_j)}(\tilde{\partial}_o)\tau(x)\right)_{1\le i, j\le2n}\alpha_1^{\lambda_1}\cdots\alpha_{2n}^{\lambda_{2n}}.
\non
\ena
Comparing the coefficient of $\alpha_1^{\lambda_1}\cdots\alpha_{2n}^{\lambda_{2n}}$, $\lambda_1>\cdots>\lambda_{2n}$, it follows that 
\bea
&&
Q_{(\lambda_1,\cdots,\lambda_{2n})}(\tilde{\partial}_o)\tau(x)=\tau(x)^{-n+1}\Pf\left(Q_{(\lambda_i,\lambda_j)}(\tilde{\partial}_o)\tau(x)\right)_{1\le i, j\le2n}
\label{eq:73}
\ena
We expand $\tau(x)$ as in (\ref{eq:51}), the coefficient $\xi_{\lambda}$ becomes (\ref{eq:111}).
Setting $x=0$ in (\ref{eq:73}) and using (\ref{eq:111}) we have (\ref{eq:52}).

\vskip5mm
Conversely we show that $\tau(x)$ given by (\ref{eq:51}) is a solution of the BKP hierarchy if $\{\xi_\lambda\}$ satisfy (\ref{eq:52}).
By (\ref{eq:51}), the coefficients $\{\xi_\lambda\}$ are defined as (\ref{eq:111}).
 

Substitute (\ref{eq:111}) to (\ref{eq:52}) and take the generating function, we have
\bea
&&
\sum_{\lambda_i\ge0}Q_{(\lambda_1,\cdots,\lambda_{2n})}(\tilde{\partial}_o)\tau(x)|_{x=0}\alpha_1^{\lambda_1}\cdots\alpha_{2n}^{\lambda_{2n}}
\non
\\
&&
=\sum_{\lambda_i\ge0}\Pf\left(Q_{(\lambda_i,\lambda_j)}(\tilde{\partial}_o)\tau(x)|_{x=0}\right)_{1\le i,j\le 2n}\alpha_1^{\lambda_1}\cdots\alpha_{2n}^{\lambda_{2n}}.
\label{eq:68}
\ena
Using (\ref{eq:63}) and the case $n=1$ of (\ref{eq:63}), (\ref{eq:68}) becomes 
\bea
&&
A_{1\cdots 2n}^{-1}\tau(2\sum_{i=1}^{2n}[\alpha_i]_o)=\Pf\left(\frac{\alpha_{ij}}{\tilde{\alpha}_{ij}}\tau(2[\alpha_i]_o+2[\alpha_j]_o)\right)_{1\le i,j\le 2n}.
\label{eq:69}
\ena
Consider the Pl\"{u}cker relation for the Pfaffians \cite{H1,O1} of (\ref{eq:69}).
Then for the odd numbers L and K, the following addition formulae hold:
\bea
&&
\sum_{l=1}^L(-1)^l A_{i_1,\cdots,i_K,j_l}^{-1}A_{j_1,\cdots,\hat{j}_l,\cdots,j_L}^{-1}\tau(2\sum_{r=1}^K[\alpha_{i_r}]_o+2[\alpha_{j_l}]_o)\tau(2\sum_{s=1,s\neq l}^L[\alpha_{j_s}]_o)
\non
\\
&&
+\sum_{k=1}^K(-1)^k A_{i_1,\cdots,\hat{i}_k,\cdots,i_K}^{-1}A_{j_1,\cdots,j_L,i_k}^{-1}\tau(2\sum_{r=1,r\neq k}^K[\alpha_{i_r}]_o)\tau(2\sum_{s=1}^L[\alpha_{j_s}]_o+2[\alpha_{i_k}]_o)=0.
\label{eq:70}
\ena

We consider the case of $L=K$ and set 
\bea
&&
j_L=i_1,\,j_{L-1}=i_2,\,j_{L-2}=i_3,\cdots,j_3=i_{L-2}.
\non
\ena
Setting $\tilde{x}=2\sum_{r=1}^{L-2}[\alpha_{i_r}]_o$ and replacing $(j_1,j_2,i_{L-1},i_L)$ by $(1,2,L-1,L)$, (\ref{eq:70}) becomes
\bea
&&
\frac{\alpha_{L-1,1}}{\tilde{\alpha}_{L-1,1}}\frac{\alpha_{L,1}}{\tilde{\alpha}_{L,1}}\frac{\alpha_{L-1,L}}{\tilde{\alpha}_{L-1,L}}\tau(\tilde{x}-2[\alpha_{L-1}]_o+2[\alpha_L]_o+2[\alpha_1]_o)\tau(\tilde{x}+2[\alpha_2]_o)
\non
\\
&&
-\frac{\alpha_{L-1,2}}{\tilde{\alpha}_{L-1,2}}\frac{\alpha_{L,2}}{\tilde{\alpha}_{L,2}}\frac{\alpha_{L-1,L}}{\tilde{\alpha}_{L-1,L}}\tau(\tilde{x}-2[\alpha_{L-1}]_o+2[\alpha_L]_o+2[\alpha_2]_o)\tau(\tilde{x}+2[\alpha_1]_o)
\non
\\
&&
+\frac{\alpha_{1,2}}{\tilde{\alpha}_{1,2}}\frac{\alpha_{1,L-1}}{\tilde{\alpha}_{1,L-1}}\frac{\alpha_{2,L-1}}{\tilde{\alpha}_{2,L-1}}\tau(\tilde{x}+2[\alpha_L]_o)\tau(\tilde{x}+2[\alpha_1]_o+2[\alpha_2]_o-2[\alpha_{L-1}]_o)
\non
\\
&&
-\frac{\alpha_{1,2}}{\tilde{\alpha}_{1,2}}\frac{\alpha_{1,L}}{\tilde{\alpha}_{1,L}}\frac{\alpha_{2,L}}{\tilde{\alpha}_{2,L}}\tau(\tilde{x}-2[\alpha_{L-1}]_o)\tau(\tilde{x}+2[\alpha_L]_o+2[\alpha_1]_o+2[\alpha_2]_o)=0.
\label{eq:71}
\ena
Equation (\ref{eq:71}) is valid if $\tilde{x}$ is replaced by an indeterminate $x$.
Shift $x$ to $x+2[\alpha_{L-1}]_o$, then (\ref{eq:71}) becomes 
\bea
&&
A_{1,2,{L-1},L}^{-1}\tau(x)\tau(x+2[\alpha_1]_o+2[\alpha_2]_o+2[\alpha_L]_o+2[\alpha_{L-1}]_o)
\non
\\
&&
=\frac{\alpha_{1,L}}{\tilde{\alpha}_{1,L}}\frac{\alpha_{2,L-1}}{\tilde{\alpha}_{2,L-1}}\tau(x+2[\alpha_1]_o+2[\alpha_L]_o)\tau(x+2[\alpha_2]_o+2[\alpha_{L-1}]_o)
\non
\\
&&
-\frac{\alpha_{2,L}}{\tilde{\alpha}_{2,L}}\frac{\alpha_{1,L-1}}{\tilde{\alpha}_{1,L-1}}\tau(x+2[\alpha_2]_o+2[\alpha_L]_o)\tau(x+2[\alpha_1]_o+2[\alpha_{L-1}]_o)
\non
\\
&&
+\frac{\alpha_{L-1,L}}{\tilde{\alpha}_{L-1,L}}\frac{\alpha_{1,2}}{\tilde{\alpha}_{1,2}}\tau(x+2[\alpha_{L-1}]_o+2[\alpha_{L}]_o)\tau(x+2[\alpha_{1}]_o+2[\alpha_{2}]_o).
\label{eq:72}
\ena
Setting $\alpha_L=0$ we have the four term equation (\ref{eq:401}).
By theorem 3 $\tau(x)$ is a solution of the BKP hierarchy.

\renewcommand{\theequation}{A.\arabic{equation}}
\setcounter{equation}{0}
\appendix
\section{The free fermions}
In the appendices we summarize necessary facts on fermions and the boson-fermion correspondence following \cite{DJKM1}.
Let $\psi_n$ and $\psi_n^*$ satisfy the following anti-commutation relations:
\bea
&&
[\psi_n,\psi_m]_+=[\psi_n^*,\psi_m^*]_+=0,
\non
\\
&&
[\psi_n,\psi_m^*]_+=\delta_{nm}.
\non
\ena
The vacuum state $|0\rangle$ and the dual vacuum state $\langle0|$ have the properties
\bea
&&
\psi_n|0\rangle=0,\,\,\,n<0,\,\,\,\,\,\psi^*_n|0\rangle=0,\,\,\,n\ge 0,
\non
\\
&&
\langle0|\psi_n=0,\,\,\,n\ge0,\,\,\,\,\,\langle0|\psi^*_n=0,\,\,\,n<0.
\non
\ena
We use the generating series of free fermionic operators
\bea
&&
\psi(z)=\sum_{k\in\mathbb{Z}}\psi_k z^k,\,\,\,\,\,\psi^*(z)=\sum_{k\in\mathbb{Z}}\psi_k^* z^{-k}.
\non
\ena
Let $H(x)$ be defined by
\bea
&&
H(x)=\sum_{l=1}^{\infty}\sum_{n\in\mathbb{Z}}x_l \psi_n \psi_{n+l}^*.
\label{eq:200}
\ena
Then the boson-fermion correspondence is valid:
\bea
&&
\langle m|e^{H(x)}\psi(k)=k^{m-1}X(k)\langle m-1|e^{H(x)},
\non
\\
&&
\langle m|e^{H(x)}\psi^*(k)=k^{-m}X^*(k)\langle m+1|e^{H(x)}.
\non
\ena

\renewcommand{\theequation}{B.\arabic{equation}}
\setcounter{equation}{0}
\section{The neutral fermions}
Let us consider $\phi_n$ satisfying 
\bea
&&
[\phi_m,\phi_n]_+=(-1)^m\delta_{m,-n}.
\non
\ena
We have the properties of the vacuum state and the dual vacuum state:
\bea
&&
\phi_n|0\rangle=0,\,\,\,n<0,
\non
\\
&&
\langle0|\phi_n=0,\,\,\,n>0.
\non
\ena
If $n=0$, we have $\phi_0^2=1/2$.\\
Set $H_B(x)$ by
\bea
&&
H_B(x)=\frac{1}{2}\sum_{l:odd}\sum_{n\in\mathbb{Z}}(-1)^{n+1}x_l\phi_n\phi_{-n-l}.
\label{eq:201}
\ena
The following the boson-fermion correspondence is valid:
\bea
&&
\langle0|\phi_0e^{H_B(x)}\phi(k)=2^{-1}X_B(k)\langle0|e^{H_B(x)},
\non
\\
&&
\langle0|e^{H_B(x)}\phi(k)=X_B(k)\langle0|\phi_0e^{H_B(x)}.
\non
\ena 

\renewcommand{\theequation}{C.\arabic{equation}}
\setcounter{equation}{0}
\section{The proof of (\ref{eq:51})}
The expansion (\ref{eq:51}) can be proved easily.
For $\alpha=(\alpha_1,\alpha_2,\cdots)$ and $\beta=(\beta_1,\beta_2,\cdots)$ we have
\bea
&&
\prod_{i,j}\frac{1+\alpha_i \beta_j}{1-\alpha_i \beta_j}=\sum_{\lambda:\text{strict}}2^{-l(\lambda)}Q^{sym}_{\lambda}(\alpha)Q^{sym}_{\lambda}(\beta)
\non
\ena
from \cite{Mac} p.255 (8.13).
The left hand side becomes
\bea
&&
\prod_{i,j}\frac{1+\alpha_i \beta_j}{1-\alpha_i \beta_j}=e^{\sum_{i,j}(\log(1+\alpha_i \beta_j)-\log(1-\alpha_i \beta_j))}
\non
\\
&&
\hskip2cm=e^{\sum_{i,j}\sum_{k=1}^{\infty}\left(\frac{(\alpha_i \beta_j)^k}{k}-\frac{(-\alpha_i \beta_j)^k}{k}\right)}
\non
\\
&&
\hskip2cm=e^{2\sum_{k:\text{odd}}k\sum_{i}\frac{\alpha_i^k}{k}\sum_{j}\frac{\beta_j^k}{k}}.
\non
\ena
Set
\bea
&&
x_k=\sum_{i}\frac{\alpha_i^k}{k},\hskip5mm y_k=\sum_{j}\frac{\beta_i^k}{k}.
\non
\ena
Then we have
\bea
&&
e^{2\sum_{k:odd}k x_k y_k}=\sum_{\lambda:\text{strict}}2^{-l(\lambda)}Q_{\lambda}(x)Q_{\lambda}(y).
\non
\ena
Replace $x_k$ by $x_k/2$ and $y_k$ by $\tilde{\partial}_{y_k}=\partial_{y_k}/k$.
We apply it to $f(y)$ and set $y=0$.
Then we get
\bea
&&
f(x)=\sum_{\lambda:\text{strict}}2^{-l(\lambda)}Q_{\lambda}\left(\frac{x}{2}\right)\left(Q_{\lambda}(\tilde{\partial}_y)f(y)\right)|_{y=0}.
\non
\ena
Setting $\xi_{\lambda}=\left(Q_{\lambda}(\tilde{\partial}_y)f(y)\right)|_{y=0}$, we obtain (\ref{eq:51}).

\begin{flushleft}{\Large {\bf Acknowledgments}}\end{flushleft}
I would like to thank Takashi Takebe for several comments.
I also thank Hirofumi Yamada for his interest in my work.
Finally I am deeply grateful to Atsushi Nakayashiki for much advice.
This work was supported by Japanese Association of University Woman JAUW.

\end{document}